# Ridesharing Services and Car-Seats: Technological Perceptions and Usage Patterns

Subasish Das, Ph.D.
Texas A&M Transportation Institute
Bryan, TX-77807
Email: s-das@tti.tamu.edu

**ABSTRACT**
Children are one of the most vulnerable groups in traffic crashes. Child safety seats (CSSs) can decrease the severity of crash outcomes for children. The usage of CSSs has significantly improved in the U.S. over the last 40 years, but it is anticipated that the usage of CSSs in popular ridesharing services (RSSs), such as Uber and Lyft, is not widespread. This paper used a publicly available nationwide internet survey that was designed to gain an understanding of riders' and drivers' perception toward child passenger safety in regard to technological perception on RSSs. This study performed a rigorous exploratory data analysis to identify the key psychological insights of the survey participants. Additionally, a recently developed dimension-reduction method has been applied to understand the co-occurrence patterns of the responses to gain intuitive insights. It is found that urban-dwelling parents with higher education degrees eventually use RSSs often due to their familiarity of the technological advantages. On the other hand, non-urban and moderately educated parents and guardians are dismissive in using RSSs while having kids with them to ride due to less trust on the technology. The findings of this study can be applied to law enforcement and policy development to promote new safety legislation, safety education design, enhancement of technology education in the non-urban areas, and the mandatory usage of CSSs in RSSs.

## 1. INTRODUCTION

Child safety is an important issue in the traffic safety field. Research has demonstrated that child safety seats (CSSs) significantly decrease injury and death in crashes when the restraints are properly implemented. Despite the safety benefits, it is anticipated that the usage of CSSs in taxicabs and other ridesharing services (RSSs), such as Uber and Lyft, is significantly lower than in personal vehicles. RSSs bring up the question of child safety on the road and how rideshare companies, rideshare drivers, and parents/guardians manage the proper usage of CSSs.

The Pew Research Center conducted an online survey in 2016. The findings show that only 30% of users who were parents and only 34% of RSS users overall believed that RSSs are a safe method of transportation for children (Smith, 2016). As RSSs are becoming popular in many urban cities, one potential solution is to mandate CSS in RSS trips. However, a regulation like this requires additional investigation.

This paper used a publicly available nationwide internet survey that was designed to gain an understanding of the riders' and drivers' perception toward CSSs in regard to RSSs (Owens et al., 2019). This survey was designed and conducted by Owens et al. (2019), which aimed to explore the current state of CSSs with respect to RSSs. The findings show that there is a general deficiency of understanding and consciousness about the problems related to the safe transportation of young children while using RSSs. The current study extended the analysis conducted by Owens et al. (2019) by using an applied and innovation dimension reduction method known as Taxicab Correspondence Analysis (TCA).

It is usually anticipated that the response patterns would vary among the parents/guardians based on the experience and willingness of using RSSs while young children and kids are with them. The objective of this study is to answer the following two key research questions:

- Are the responses different for the group that has used RSSs with children (kids) and the group that has never used RSSs with kids? Do education, residence location, and technology perception play role?
- If there are significantly different responses for certain variables, what are the response patterns for both of these groups?

This study designed the approach to answer these two key questions. As survey data are categorical in nature, it is important to identify the suitable method to explore the unknown trends. TCA is an excellent survey analysis tool that can provide multiple contexts from a low dimensional space by identifying the closeness of the question responses in the form of the clusters. The conclusions of this important study can assist policy makers in understanding the problems associated with CSS usages in RSSs to make data-driven decisions.

## 2. LITERATURE REVIEW

There are a handful of studies that focused on the safe usage of the CSSs. As developed countries like the U.S. have established a wider usage of CSSs in privately owned cars, the brief literature review is based on the studies in the U.S. and outside of the U.S.

### 2.1. Studies Outside of the U.S.

Desapriya et al. (2008) explored the legislative impact on fatalities in Japan during 1994-2005. Kakefuda et al. (2008) assessed associations between the usage of the CSSs and mindsets among Japanese mothers. For short ride, they found some key contributing factors: child resistance to use CSSs for short period; hassles associated with the usage of CSSs; and less eagerness of the mother. Kulanthayan et al. (2010) determined the CSS usage rates in Malaysia to evaluate driver behaviours that are associated with CSS usage. Results showed that 27% of the drivers used CSS at least once during the time of the survey. Using data from Russia, Ma et al. (2012) examined the usage of CSSs and associated insights, attitudes, and perceptions of the safety of the usage of CSSs. Bromfield and Mahmoud (2017) conducted a United Arab Emirates (UAE) based preliminary investigation on the usages of CSSs among UAE citizen parents (n=366 citizens with at least one child younger than age 13 years). The findings call for additional studies on CSS usage in the UAE. Moradi et al. *(2017)* determined the prevalence of CSS use and the elements influencing its use in Tehran, Iran. Nazif-Munoz and Nikolic *(2018)* examined the short- and long-term impacts of Serbia's 2009 update of the legislation on the usage of CSS. The results show that the new law showed both short term (for children aged 0–3 years old) and long term (for children aged 4–12 years old) safety effectiveness. Using data from Iran, Moradi et al. *(2019)* determined the prevalence of child safety seat use in vehicles to determine the influential factors.

### 2.2. Studies Inside the U.S.

Farmer et al. (2009) examined the relationship between booster seat legislatures and fatalities among children in vehicle frontal crashes. The results show that states with a booster seat law showed high safety performance on child safety compared to the states without such law. Brown et al. (2009) trained and monitored twenty-seven adult participants in three stages of CSS installation. Thoreson et al. (2009) compared childcare center–based booster seat education and distribution with no intervention. Kroeker et al. (2015) analyzed the CSS inspection forms for two SafeKids Worldwide Coalitions in Michigan. The retrospective, cross-section analysis showed that 10% of car seat inspections are for booster seats and 50% are for rear-facing car seats. Billie et al. (2016) conducted a study to increase the use of safety seats among tribal communities to improve the safety of the children. Asbridge et al. (2018) performed a thorough systematic review and meta-analysis of observational studies that focused on CSSs to decrease injury severity from crashes among child passengers (age:4-8 years old). Ojo (2018), performing an unhindered observational survey guided by the Theory of Planned Behaviour, inspected seat belt use by drivers and accompanied child during school related trips using data from Ghana.

Olsen et al. (2010) examined the association between the restraint usage of the driver and child emergency department (ED) assessment following a traffic collision. Sun et al. (2010) examined the influence of the New York State upgraded child restraint law (UCRL) implemented in 2005 and the following crash injury rate among four to six years old. Jeffrey et al. (2016) investigated the descriptive norms in health and safety appeals towards booster seat usage.

The research findings show there are almost no studies, except Owens et al. (2019), which focused on the usage of CSSs in RSSs. The current study, which is based on the survey conducted by Owens et al. (2019), performed a rigorous study to explore the insights from the response patterns.

## 3. METHODOLOGY

### 3.1. Survey Data

This study collected the open-source data available to conduct the study (Owens et al., 2019). This study was conducted by the Texas A&M Transportation Institute (TTI) and Virginia Tech Transportation Institute (VTTI). The survey sample comprised approximately 1,200 participants. The participants were all parents that have children up to the age of five, which is the age at which the transition from harness child seats to booster seats is made for many children. The locations are selected based on the availability of UberFAMILY car seat program. The participants were randomly selected from those locations.

The survey questionnaire can be found in Owens et al. (2019) report. There are five questions about the participants' personal information, including their gender, education level, age group, where they live, and how many children they care for. The rest of the questions regard information about their travel habits with children and, specifically, their use of RSSs with children. The participants answered questions that applied to their situation; for example, if they answered that they have never used RSSs with children, then they were asked why they have not used an RSS with children (aged under 5) under their supervision. If they have used an RSS with children, then they were also asked questions about what safety precautions were taken and what their experience was like. There are also several questions regarding child safety laws and the legal responsibilities of the parent and the driver.

### 3.2. Child Safety Laws in Different States

Figure 1 illustrates four maps of the United States to provide a visual representation of child safety laws in different states. The map in the upper left corner shows which states have primary enforcement of child safety laws. The states shown in a brown color do have primary enforcement of child safety laws, while the states shown in blue do not. Thirty-six of the fifty states have primary enforcement of child safety laws. The fourteen states that do not have primary enforcement of child safety laws include (in alphabetical order) Arizona, Colorado, Idaho, Massachusetts, Montana, Nebraska, Nevada, New Hampshire, North Dakota, Ohio, South Dakota, Vermont, Virginia, and Wyoming. The upper right corner of Figure 1 shows an illustration of the age at which child safety laws are in effect in various states. Most states have laws for children less than seven or eight years old. Twenty states have child safety laws in effect for children under the age of seven years old, and seventeen states have laws that for children under the age of eight years old. The bottom left corner of Figure 1 contains a map that illustrates the liability of the driver for children's safety. The map shows that the driver is liable in only fourteen of the states. The bottom right corner of Figure 1, however, shows that the ridesharing or taxi company is liable for children's safety in thirty-three of the states.

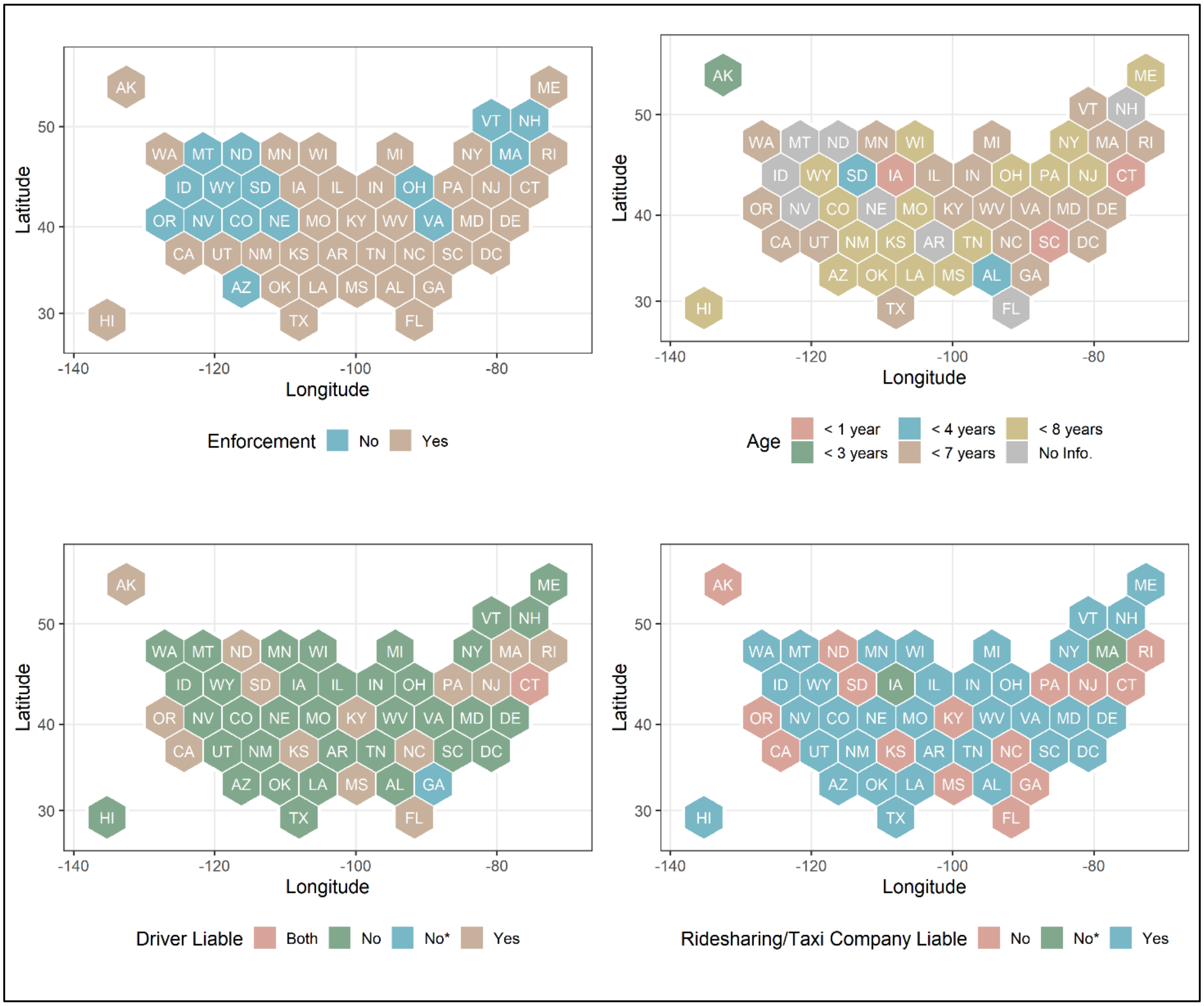

**Figure 1. Regulation on CSSs by the U.S. states**

### 3.3. Participant Characteristics

Figure 2 illustrates the location, age, educational degree, and gender of the participants who have used RSSs with kids or who have never used RSSs with kids. The green bars represent the percentage of participants in that demographic that have used RSSs with children present, while the red bars represent those that have never used RSSs with children present. One of the demographics studied was the location of the participants, found in Question 36. The highest percentage (40%) of people who have never used RSSs with children present were located in a suburb. The highest percentage of those who have used RSSs with children present (31%) were also located in the suburbs. Although, it was a smaller percentage than those who have. The opposite is true for participants in large cities where a higher percentage of people who have used RSSs with children are located (27%) and a lower percentage of people who have never used RSSs with children are located (15%). Question 37 studied the age of the participants. Figure 2 shows that a majority of participants who have used RSSs with children present are in the age group of 22-30 years or 31-45 years old (45% for both demographics). The majority of participants who have never used RSSs with children present are in the age group of 22-30 years or 31-45 years old;

however, the percentages are low: 35% and 52%, respectively. Another demographic studied was the degree level of participants. Participants having higher education (advanced degree or college degree) are more likely to rideshare with children present. The opposite trend is visible for those who have some college or high school level education. The final demographic studied was the gender of the respondents. There was no noteworthy difference between the percentage of females who have used RSSs with kids (61%) and the females who have never used RSSs with kids (62%). The same is true for males who have used RSSs with kids (27%) and the males who have never used RSSs with kids (26%).

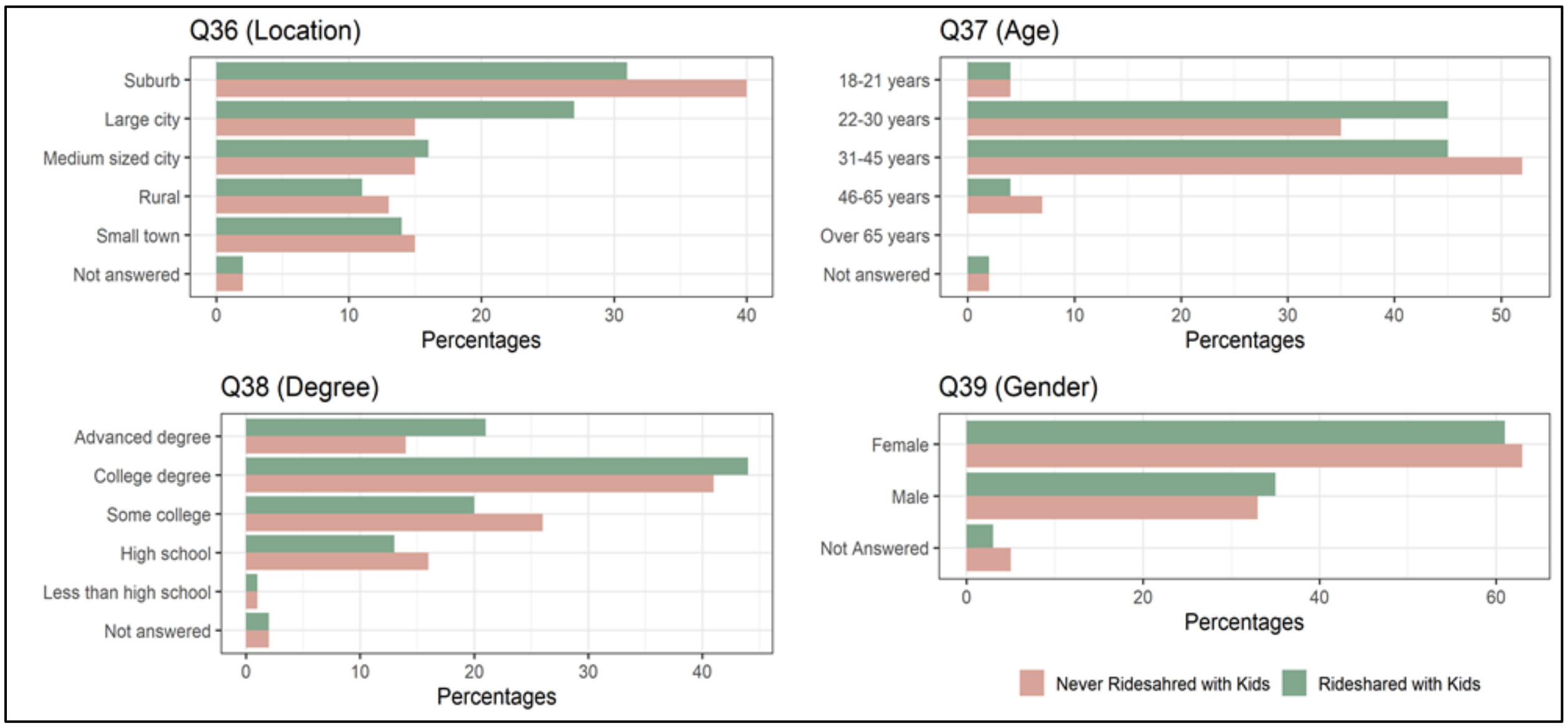

**Figure 2. Traits of the participants**

### 3.4. Comparison between Two Groups

The common significant questions (n=24) are selected for comparing the groups: 1) group of respondents who used RSSs with kids, and 2) group of respondents who never used RSSs with kids. Table 1 lists the comparison test measures for the selected questions. The research team used open source software R to conduct the analysis (version 3.6.0). For the descriptive comparisons, R package 'compareGroups' (Salvadore, 2020) was used. This study defined statistical significance as p-value < 0.05. Nine questions are not statistically significant at a 95% confidence interval. It is found that there is no statistically significant difference between groups when they are asked about their knowledge of state laws regarding car seat guidelines. The majority of the Question 28 queries are not statistically significant. The broader perspective of Question 28 helps to understand the importance of the key factors that the respondents consider when deciding whether to use an RSS with children (aged under 5). The identified factors in the question are 1) pre-installed child seat available, 2) rear seat space, 3) driver's driving record, 4) type of vehicle, 5) convenience of child seat installation in a vehicle, 6) vehicle cleanliness, and 7) information abundance on child seat laws and regulations. Both groups differ in responses for only one query (driver's driving record) from the set of queries. Gender is also found as not statistically significant. The comparison of the response patterns answers the research question 1. The results show that these two groups have different views on several major issues when they consider using an RSS with young children or kids.

**Table 1. Comparison of the responses provided by two main groups.**

| Question | Never Rideshared with Kids (N=409) | Rideshared with Kids (N=768) | p-overall |
|---|---|---|---|
| ***17: In what situations and how often have you used a ride share service such as Uber or Lyft without a child or children?*** | | | |
| *17a: For local travel while on an out-of-town trip* | | | **<0.001** |
| Have used only once | 54 (13.2%) | 120 (15.6%) | |
| Used a few times (2 -5 times) | 122 (29.8%) | 202 (26.3%) | |
| Used often (6-10 times) | 55 (13.4%) | 196 (25.5%) | |
| Used regularly (more than 10 times) | 38 (9.29%) | 150 (19.5%) | |
| Never | 140 (34.2%) | 100 (13.0%) | |
| *17b: For routine local travel where I live* | | | **<0.001** |
| Have used only once | 46 (11.2%) | 125 (16.4%) | |
| Used a few times (2 -5 times) | 67 (16.4%) | 209 (27.4%) | |
| Used often (6-10 times) | 54 (13.2%) | 164 (21.5%) | |
| Used regularly (more than 10 times) | 19 (4.65%) | 101 (13.2%) | |
| Never | 223 (54.5%) | 165 (21.6%) | |
| *17c: For non-routine local travel where I live* | | | **<0.001** |
| Have used only once | 58 (14.3%) | 132 (17.3%) | |
| Used a few times (2 -5 times) | 63 (15.5%) | 201 (26.3%) | |
| Used often (6-10 times) | 40 (9.83%) | 166 (21.8%) | |
| Used regularly (more than 10 times) | 18 (4.42%) | 105 (13.8%) | |
| Never | 228 (56.0%) | 159 (20.8%) | |
| *17d:To make an out-of-town trip* | | | **<0.001** |
| Have used only once | 32 (7.84%) | 105 (13.9%) | |
| Used a few times (2 -5 times) | 43 (10.5%) | 139 (18.3%) | |
| Used often (6-10 times) | 27 (6.62%) | 146 (19.3%) | |
| Used regularly (more than 10 times) | 21 (5.15%) | 94 (12.4%) | |
| Never | 285 (69.9%) | 274 (36.1%) | |
| ***18: In what situations and how often have you used a taxi when transporting a child or children under your care?*** | | | |
| *18a:For local travel while on an out-of-town trip* | | | **<0.001** |
| Have used only once | 43 (10.6%) | 132 (17.2%) | |
| Used a few times (2 -5) | 66 (16.2%) | 174 (22.7%) | |
| Used often (6-10 times) | 23 (5.65%) | 110 (14.4%) | |
| Used regularly (more than 10 times) | 17 (4.18%) | 104 (13.6%) | |
| Never | 258 (63.4%) | 246 (32.1%) | |
| *18b: For routine local travel where I live* | | | **<0.001** |
| Have used only once | 41 (10.1%) | 107 (14.0%) | |
| Used a few times (2 -5) | 33 (8.11%) | 163 (21.3%) | |
| Used often (6-10 times) | 23 (5.65%) | 129 (16.9%) | |
| Used regularly (more than 10 times) | 9 (2.21%) | 68 (8.89%) | |
| Never | 301 (74.0%) | 298 (39.0%) | |
| *18c: For non-routine local travel where I live* | | | **<0.001** |
| Have used only once | 33 (8.09%) | 125 (16.3%) | |
| Used a few times (2 -5) | 42 (10.3%) | 143 (18.7%) | |
| Used often (6-10 times) | 18 (4.41%) | 117 (15.3%) | |
| Used regularly (more than 10 times) | 11 (2.70%) | 76 (9.93%) | |
| Never | 304 (74.5%) | 304 (39.7%) | |
| *18d: To make an out-of-town trip* | | | **<0.001** |
| Have used only once | 26 (6.42%) | 93 (12.3%) | |
| Used a few times(2 -5) | 23 (5.68%) | 113 (14.9%) | |
| Used often (6-10 times) | 21 (5.19%) | 115 (15.2%) | |
| Used regularly (more than 10 times) | 17 (4.20%) | 76 (10.0%) | |
| Never | 318 (78.5%) | 362 (47.7%) | |
| *Q21: How familiar are you with the recommended child seat guidelines for the age or ages of your children?* | | | 0.591 |
| Very | 264 (64.9%) | 495 (64.7%) | |
| Somewhat | 124 (30.5%) | 242 (31.6%) | |

| Question | Never Rideshared with Kids (N=409) | Rideshared with Kids (N=768) | p-overall |
|---|---|---|---|
| A little | 14 (3.44%) | 24 (3.14%) | |
| Not at all | 5 (1.23%) | 4 (0.52%) | |
| *Q22: How familiar are you with your state's laws for child safety seat use for the age or ages of your children?* | | | 0.886 |
| Very | 235 (58.8%) | 456 (60.7%) | |
| Somewhat | 130 (32.5%) | 234 (31.2%) | |
| A little | 28 (7.00%) | 51 (6.79%) | |
| Not at all | 7 (1.75%) | 10 (1.33%) | |
| *Q24: How confident are you that you have followed state laws when using taxi services with children under 5 years old?* | | | **0.001** |
| Very confident | 86 (46.7%) | 270 (48.0%) | |
| Somewhat confident | 56 (30.4%) | 187 (33.2%) | |
| Not very confident | 16 (8.70%) | 69 (12.3%) | |
| Not at all confident | 16 (8.70%) | 34 (6.04%) | |
| Have not used taxis with children under 5 years old | 10 (5.43%) | 3 (0.53%) | |
| *Q25: Who is legally responsible for correct child safety seat use for children under 5 years old as passengers in ride-share services such as Uber or Lyft?* | | | **<0.001** |
| The parent, guardian, or caregiver | 221 (54.4%) | 397 (52.1%) | |
| The driver | 88 (21.7%) | 179 (23.5%) | |
| The ride share company | 15 (3.69%) | 63 (8.27%) | |
| Depends on the circumstances | 7 (1.72%) | 46 (6.04%) | |
| Depends on the State | 17 (4.19%) | 18 (2.36%) | |
| Not sure | 58 (14.3%) | 59 (7.74%) | |
| *Q26: Who is legally responsible for correct child safety seat use for children under 5 years old as passengers in taxis?* | | | **<0.001** |
| The parent, guardian, or caregiver | 222 (54.7%) | 391 (51.7%) | |
| The driver | 93 (22.9%) | 175 (23.1%) | |
| Depends on the circumstances | 13 (3.20%) | 85 (11.2%) | |
| Depends on the State | 20 (4.93%) | 35 (4.62%) | |
| Not sure | 58 (14.3%) | 71 (9.38%) | |
| ***28: How important are each of the factors below in your decisions about using a ride-share service with a child or children under the age of 5?*** | | | |
| *28a: Pre-installed child seat available* | | | 0.033 |
| Very important | 197 (52.5%) | 327 (43.8%) | |
| Somewhat important | 94 (25.1%) | 210 (28.1%) | |
| Somewhat unimportant | 47 (12.5%) | 106 (14.2%) | |
| Not at all important | 37 (9.87%) | 104 (13.9%) | |
| *28b: Rear seat space* | | | 0.429 |
| Very important | 226 (59.8%) | 423 (56.0%) | |
| Somewhat important | 99 (26.2%) | 220 (29.1%) | |
| Somewhat unimportant | 36 (9.52%) | 86 (11.4%) | |
| Not at all important | 17 (4.50%) | 27 (3.57%) | |
| *28c: Driver safety record* | | | **0.002** |
| Very important | 292 (76.0%) | 490 (65.5%) | |
| Somewhat important | 57 (14.8%) | 158 (21.1%) | |
| Somewhat unimportant | 22 (5.73%) | 75 (10.0%) | |
| Not at all important | 13 (3.39%) | 25 (3.34%) | |
| *28d: Type of vehicle* | | | 0.477 |
| Very important | 162 (42.7%) | 291 (38.9%) | |
| Somewhat important | 135 (35.6%) | 273 (36.4%) | |
| Somewhat unimportant | 54 (14.2%) | 130 (17.4%) | |
| Not at all important | 28 (7.39%) | 55 (7.34%) | |
| *28e: Convenience of child seat installation in vehicle* | | | 0.043 |
| Very important | 209 (55.7%) | 349 (47.2%) | |
| Somewhat important | 102 (27.2%) | 240 (32.4%) | |
| Somewhat unimportant | 49 (13.1%) | 106 (14.3%) | |
| Not at all important | 15 (4.00%) | 45 (6.08%) | |
| *28f: Vehicle cleanliness* | | | 0.63 |

| Question | Never Rideshared with Kids (N=409) | Rideshared with Kids (N=768) | p-overall |
|---|---|---|---|
| Very important | 210 (55.1%) | 389 (51.7%) | |
| Somewhat important | 121 (31.8%) | 246 (32.7%) | |
| Somewhat unimportant | 36 (9.45%) | 87 (11.6%) | |
| Not at all important | 14 (3.67%) | 30 (3.99%) | |
| *28g: Information abundance on child seat laws and regulations* | | | 0.029 |
| Very important | 159 (43.2%) | 282 (38.5%) | |
| Somewhat important | 116 (31.5%) | 215 (29.4%) | |
| Somewhat unimportant | 57 (15.5%) | 170 (23.2%) | |
| Not at all important | 36 (9.78%) | 65 (8.88%) | |
| *Q36: Location where you live?* | | | **0.001** |
| Suburb | 165 (40.8%) | 239 (31.3%) | |
| Large city | 62 (15.3%) | 209 (27.4%) | |
| Medium sized city | 60 (14.9%) | 123 (16.1%) | |
| Rural | 54 (13.4%) | 85 (11.1%) | |
| Small town | 63 (15.6%) | 108 (14.1%) | |
| *Q37: Which age group are you in?* | | | **0.004** |
| 18-21 years | 18 (4.46%) | 33 (4.32%) | |
| 22-30 years | 143 (35.4%) | 351 (45.9%) | |
| 31-45 years | 216 (53.5%) | 347 (45.4%) | |
| 46-65 years | 27 (6.68%) | 32 (4.19%) | |
| Over 65 years | 0 (0.00%) | 1 (0.13%) | |
| *Q38: What is your education level?* | | | **0.001** |
| Advanced degree | 59 (14.6%) | 161 (21.1%) | |
| College degree | 170 (42.0%) | 340 (44.6%) | |
| Some college | 106 (26.2%) | 153 (20.1%) | |
| High school | 67 (16.5%) | 103 (13.5%) | |
| Less than high school | 3 (0.74%) | 6 (0.79%) | |
| *Q39: What is your gender?* | | | 0.655 |
| Female | 258 (65.6%) | 476 (63.3%) | |
| Male | 135 (34.4%) | 275 (36.6%) | |

## 3.5. Criteria for using RSSs with Kids

Question 17 and Question 18 are designed to collect the perspective of the users on the usage of RSSs without children and the usage of taxis with children. Each question has four scenarios:

- For local travel for an out-of-town trip
- For routine local travel where I live
- For non-routine local travel in the residence location
- To make an out-of-town trip

The participants need to choose one from the five multiple choices (never, have used once, used 2-3 times, used 6-10 times, and used more than 10 times) for each of the scenarios from these two questions. Figure 3 shows the response patterns of these two questions for both groups: participants who have used RSSs with children and participants who have never used RSSs with children. One clear distinction is that majority of the participants who have never used RSSs with children chose the response 'never' for eight scenarios (ranges from 35% to 100% based on scenario type). Another common finding is that the response 'never' is likely to be higher if the education is high school or less for both groups. A pattern shown in the responses of participants who have used RSSs with children present is that they were more likely to respond 'never' to question 18 than question 17, meaning they were more likely to have never used taxis for ridesharing with children than to have never used RSSs like Uber or Lyft with children present.

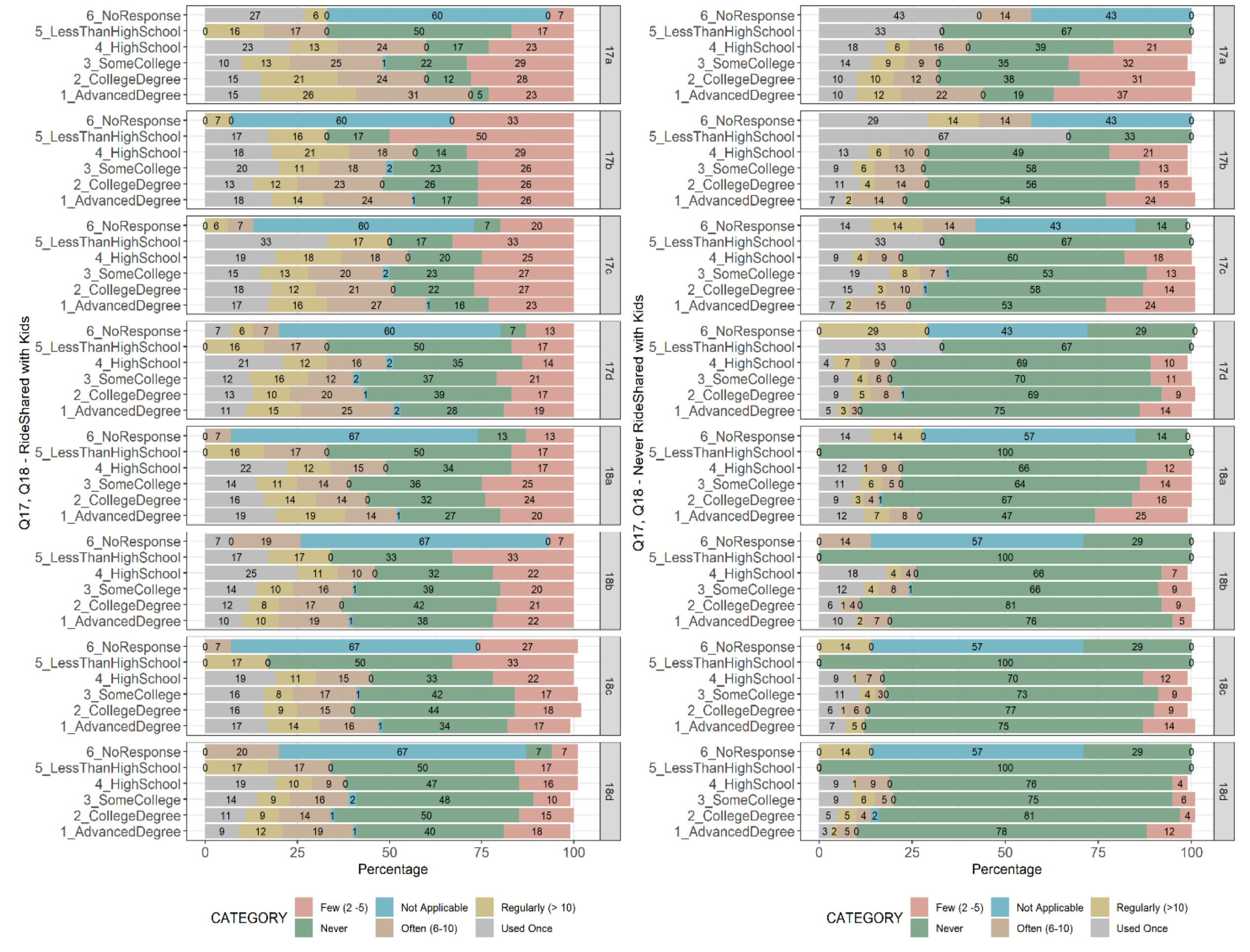

| Group 1 | Group 2 |
|---|---|

**Figure 3. Response patterns of Questions 17 and 18.**

Question 28 asked the participants about the importance of various factors when deciding to use an RSS with children (aged under 5). The factors were as follows:

- Pre-installed child seat available (choose from five options)
- Rear seat space (choose from five options)
- Driver safety record (choose from five options)
- Type of vehicle (choose from five options)
- Convenience of child seat installation in vehicle (choose from five options)
- Vehicle cleanliness (choose from five options)
- Information abundance regarding child seat laws and regulations (choose from five options)

The participants had to choose from one of five responses (not applicable, not important, somewhat unimportant, somewhat important, or important) for each factor. Figure 4 shows the response patterns of these two questions for both groups: participants who have used RSSs with children and participants who have never used RSSs with children. One major takeaway is that a majority of participants from both groups chose 'Important' for each of the seven factors, ranging from 33% to 75% depending on the factor, this excludes the participants that chose 'No Response' for the given demographic. Almost every factor also had a high percentage of the response 'Somewhat Important' for both groups of participants as well. This response for each of the factors ranged from 15% to 50% for participants that have used RSSs with children present, and it ranged from 9% to 39% for participants that have never used RSSs with children present (these percentages also exclude the participants that chose 'No Response' for the given demographic).

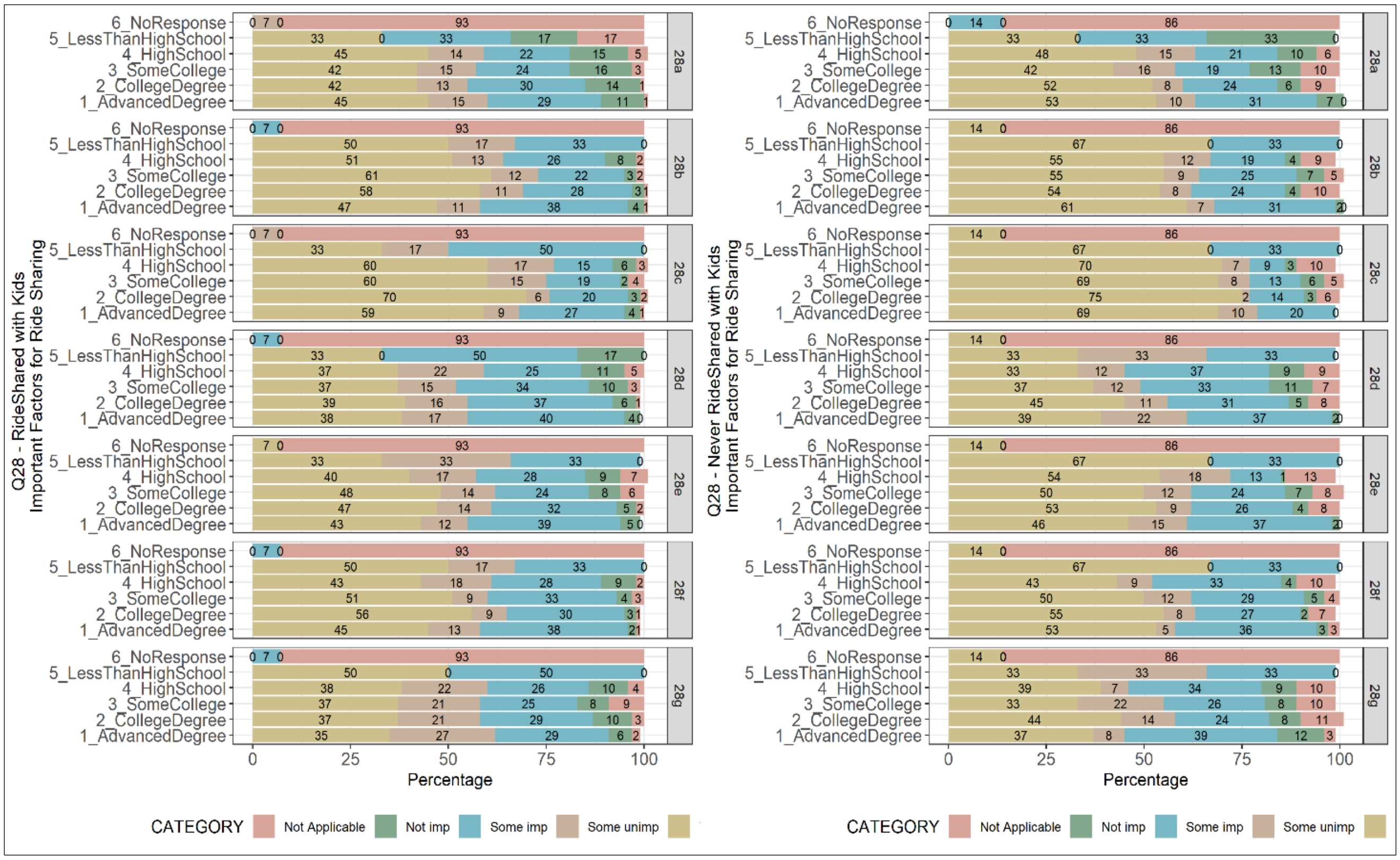

(a) Group 1 (Rideshared with kids)    (b) Group 2 (Never rideshared with kids)

**Figure 4. Response patterns of Question 28.**

## 3.6. Taxicab Correspondence Analysis

An enhanced version of correspondence analysis (CA), known as TCA, has been develop by Choulakian in a series of papers (Choulakian, 2013; Choulakian, 2006; Choulakian, 2006). Different variants of CA have been used by the transportation researchers in the recent years (Das and Sun, 2015; Das and Sun, 2016; Jalayer et al., 2018).

Euclidean distance measure is used in CA, whereas TCA uses Manhattan city block as the distance value, which is also known as taxicab distance. Let $X = (x_1, x_2, \ldots\ldots, x_n)$ and $Y = (y_1, y_2, \ldots\ldots, y_n)$ and a vector $\boldsymbol{v} = (v_1, v_2, \ldots\ldots, v_n)$ to present these two distances:

$$Euclidean\ Distance = ED(X,Y) = \sqrt{\sum_{i=1}^{n}(x_i - y_i)^2}\ \text{[with } L_2 \text{ Norm=} \|v\|_2 = \sqrt{\sum_{i=1}^{n}(v_i)^2}\text{]} \quad (1)$$

$$Taxicab\ Distance = TD(X,Y) = \sum_{i=1}^{n}|x_i - y_i| \text{ [with } L_1 \text{ Norm=}\|v\|_1 = \sum_{i=1}^{n}|v_i|\text{]} \quad (2)$$

The concepts TCA lie within singular value decomposition (SVD), more specifically taxicab singular value decomposition (TSVD). Let a real matrix $A$ is decomposed as $M\Lambda^{1/2}N'$, with $\Lambda$ the diagonal matrix of the eigenvalues of $AA', M$ the orthogonal matrix of the corresponding eigenvectors, and $N$ the matrix of eigenvectors of $A'A$ (with constraints $M'M = I\ and\ N'N = I$). Choulakian (2006) used a recurse optimization technique to evaluate the solutions of SVD. The relations between rows and columns of N are summarized by the $\chi^2$ statistics, which quantity the association departure from the scenarios of independence. TCA Analysis is defined as the Taxicab SVD of the data table $D = T - rl'$; it considers the table's profiles, respectively $R = D_r^{-1}D$ for the rows and $L = D_l^{-1}D$ for the columns.

## 4. RESULTS AND DISCUSSIONS

'TaxicabCA' (Allard and Choulakian, 2019), a package developed for R, was used to perform TCA in tis study. TCA analysis has been performed on the datasets developed for both groups (participants who have used RSSs with kids, and participants who have never used RSSs with kids). The questions are selected based on the prior comparison test measures. The analytical results produce different performance measures (for example, column score, row score, and variance explained by axes). Table 2 lists the column scores for both groups. The signs of the axis values are used in assigning the quadrants. The outcomes of Table 1 are shown in two-dimensional plots in Figure 5 and Figure 6. The clusters are developed based on the proximity of the co-ordinates associated with the responses. Only the first two axes have been considered for analysis. The variances explained by Axis 1 and Axis 2 for group 1 (participants who have used RSSs with kids) are 44.83% and 33.32%. For the other groups, the variance explanation capacities by Axis 1 and Axis 2 are 39.65% and 30.11%, respectively.

**Table 2. Column scores from separate TCA analysis on both groups.**

| Participants Rideshared with Kids | | | | | Participants Never Rideshared with Kids | | | | |
|---|---|---|---|---|---|---|---|---|---|
| **Ques.** | **Response** | **Axis1** | **Axis2** | **Quad.** | **Ques.** | **Response** | **Axis1** | **Axis2** | **Quad.** |
| 17a | > 10 times | 0.34 | -0.64 | 4th | 17a | 1 | 0.42 | 0.28 | 1st |
| 17a | 0 | -0.78 | -0.27 | 3rd | 17a | Not appli | 1.12 | 1.11 | 1st |
| 17a | 1 | 0.26 | 0.57 | 1st | 17b | > 10 times | 0.91 | 0.57 | 1st |
| 17a | 2-5 times | -0.21 | 0.46 | 2nd | 18b | > 10 times | 0.90 | 0.20 | 1st |
| 17a | 6-10 times | 0.25 | -0.25 | 4th | 18c | > 10 times | 1.12 | 0.20 | 1st |
| 17a | Not appli | -1.08 | 1.14 | 2nd | 17b | Not appli | 1.12 | 1.11 | 1st |
| 17b | > 10 times | 0.41 | -0.60 | 4th | 17c | > 10 times | 1.01 | 0.54 | 1st |
| 17b | 0 | -0.94 | -0.22 | 3rd | Q25 | Parent/Caregiver | 0.23 | 0.36 | 1st |
| 17b | 1 | 0.15 | 0.52 | 1st | Q26 | Parent/Caregiver | 0.19 | 0.41 | 1st |
| 17b | 2-5 times | 0.20 | 0.46 | 1st | Q36 | Large city | 0.25 | 0.24 | 1st |
| 17b | 6-10 times | 0.42 | -0.48 | 4th | 17c | Not appli | 0.72 | 0.68 | 1st |
| 17b | Not appli | -1.08 | 1.14 | 2nd | 17d | Not appli | 1.12 | 1.11 | 1st |
| 17c | > 10 times | 0.50 | -0.64 | 4th | 18a | Not appli | 0.72 | 1.08 | 1st |
| 17c | 0 | -0.98 | -0.30 | 3rd | 18b | Not appli | 1.12 | 1.11 | 1st |
| 17c | 1 | 0.01 | 0.59 | 1st | 18c | Not appli | 1.12 | 1.11 | 1st |
| 17c | 2-5 times | 0.15 | 0.38 | 1st | 18d | Not appli | 0.55 | 0.78 | 1st |
| 17c | 6-10 times | 0.53 | -0.33 | 4th | 28c | Not appli | 0.19 | 0.19 | 1st |
| 17c | Not appli | -1.08 | 1.14 | 2nd | Q25 | Not appli | 0.79 | 0.75 | 1st |
| 17d | > 10 times | 0.65 | -0.68 | 4th | Q26 | Not appli | 0.79 | 0.75 | 1st |
| 17d | 0 | -0.81 | -0.07 | 3rd | Q36 | Not appli | 1.12 | 1.11 | 1st |

| Participants Rideshared with Kids | | | | | Participants Never Rideshared with Kids | | | | |
|---|---|---|---|---|---|---|---|---|---|
| **Ques.** | **Response** | **Axis1** | **Axis2** | **Quad.** | **Ques.** | **Response** | **Axis1** | **Axis2** | **Quad.** |
| 17d | 1 | 0.31 | 0.54 | 1st | Q37 | Not appli | 1.12 | 1.11 | 1st |
| 17d | 2-5 times | 0.35 | 0.48 | 1st | Q38 | Not appli | 1.12 | 1.11 | 1st |
| 17d | 6-10 times | 0.68 | -0.36 | 4th | Q36 | Rural | 0.08 | 0.37 | 1st |
| 17d | Not appli | -0.88 | 0.62 | 2nd | Q36 | Small town | 0.26 | 0.26 | 1st |
| 18a | > 10 times | 0.77 | -0.78 | 4th | Q37 | 18-21 | 0.68 | 0.63 | 1st |
| 18a | 0 | -0.96 | -0.20 | 3rd | Q37 | 22-30 | 0.16 | 0.36 | 1st |
| 18a | 1 | 0.26 | 0.64 | 1st | 17d | > 10 times | 0.84 | 0.14 | 1st |
| 18a | 2-5 times | 0.28 | 0.43 | 1st | Q38 | High school | 0.38 | 0.55 | 1st |
| 18a | 6-10 times | 0.76 | -0.39 | 4th | Q38 | Some college | 0.22 | 0.43 | 1st |
| 18a | Not appli | -0.91 | 1.12 | 2nd | 17a | 0 | -0.44 | 0.35 | 2nd |
| 18b | > 10 times | 0.75 | -0.73 | 4th | 17b | 0 | -0.44 | 0.21 | 2nd |
| 18b | 0 | -0.93 | -0.16 | 3rd | 18a | 0 | -0.41 | 0.22 | 2nd |
| 18b | 1 | 0.40 | 0.69 | 1st | 18b | 0 | -0.39 | 0.10 | 2nd |
| 18b | 2-5 times | 0.60 | 0.36 | 1st | 18c | 0 | -0.37 | 0.15 | 2nd |
| 18b | 6-10 times | 0.77 | -0.38 | 4th | 18d | 0 | -0.29 | 0.13 | 2nd |
| 18b | Not appli | -1.08 | 0.98 | 2nd | 17c | 0 | -0.44 | 0.18 | 2nd |
| 18c | > 10 times | 0.84 | -0.88 | 4th | 28c | Very important | -0.17 | 0.08 | 2nd |
| 18c | 0 | -0.99 | -0.16 | 3rd | Q36 | Medium sized city | -0.15 | 0.09 | 2nd |
| 18c | 1 | 0.51 | 0.58 | 1st | Q24 | Not appli | -0.51 | 0.24 | 2nd |
| 18c | 2-5 times | 0.66 | 0.41 | 1st | Q37 | 46-65 | -0.21 | 0.28 | 2nd |
| 18c | 6-10 times | 0.80 | -0.26 | 4th | 17d | 0 | -0.32 | 0.15 | 2nd |
| 18c | Not appli | -1.08 | 1.14 | 2nd | 17a | 2-5 times | -0.08 | -0.35 | 3rd |
| 18d | > 10 times | 0.90 | -0.91 | 4th | Q25 | Depends on the State | -0.17 | -0.62 | 3rd |
| 18d | 0 | -0.80 | -0.10 | 3rd | Q25 | Driver | -0.33 | -0.53 | 3rd |
| 18d | 1 | 0.54 | 0.67 | 1st | Q25 | Not sure | -0.50 | -0.38 | 3rd |
| 18d | 2-5 times | 0.80 | 0.58 | 1st | Q25 | Ride share company | -0.21 | -0.18 | 3rd |
| 18d | 6-10 times | 0.84 | -0.32 | 4th | Q26 | Driver | -0.28 | -0.53 | 3rd |
| 18d | Not appli | -0.76 | 0.79 | 2nd | Q26 | Not sure | -0.46 | -0.51 | 3rd |
| 28c | Not appli | -0.48 | 0.75 | 2nd | Q26 | State | -0.38 | -0.39 | 3rd |
| 28c | Not at all important | 0.52 | 0.11 | 1st | Q36 | Suburb | -0.22 | -0.40 | 3rd |
| 28c | Some important | 0.32 | 0.19 | 1st | Q37 | 31-45 | -0.17 | -0.37 | 3rd |
| 28c | Some unimportant | 0.66 | 0.53 | 1st | Q38 | Advanced degree | -0.13 | -0.74 | 3rd |
| 28c | Very important | -0.20 | -0.19 | 3rd | Q38 | College degree | -0.27 | -0.27 | 3rd |
| Q24 | Never used taxis with kids | 0.92 | 0.96 | 1st | Q38 | Less than high school | -0.88 | -0.36 | 3rd |
| Q24 | Not appli | -1.04 | -0.25 | 3rd | 17a | > 10 times | 0.38 | -0.36 | 4th |
| Q24 | Not at all confident | 0.04 | 0.45 | 1st | 17a | 6-10 times | 0.54 | -0.20 | 4th |
| Q24 | Not very confident | 0.31 | 0.58 | 1st | 17d | 1 | 0.50 | -0.50 | 4th |
| Q24 | Some confident | 0.43 | 0.26 | 1st | 17d | 2-5 times | 0.61 | -0.55 | 4th |
| Q24 | Very confident | 0.43 | -0.20 | 4th | 17d | 6-10 times | 1.05 | -0.38 | 4th |
| Q25 | Circumstances | 0.27 | 0.58 | 1st | 17b | 1 | 0.43 | -0.24 | 4th |
| Q25 | Not appli | -0.83 | 0.87 | 2nd | 18a | > 10 times | 1.00 | -0.31 | 4th |
| Q25 | Not sure | -0.94 | 0.72 | 2nd | 18a | 1 | 0.70 | -0.32 | 4th |
| Q25 | Parent/Caregiver | 0.04 | -0.12 | 4th | 18a | 2-5 times | 0.52 | -0.45 | 4th |
| Q25 | Ride share company | 0.51 | 0.46 | 1st | 18a | 6-10 times | 0.95 | -0.64 | 4th |

| Participants Rideshared with Kids | | | | | Participants Never Rideshared with Kids | | | | |
|---|---|---|---|---|---|---|---|---|---|
| **Ques.** | **Response** | **Axis1** | **Axis2** | **Quad.** | **Ques.** | **Response** | **Axis1** | **Axis2** | **Quad.** |
| Q25 | State | -0.52 | 0.42 | 2nd | 17b | 2-5 times | 0.40 | -0.46 | 4th |
| Q25 | Driver | 0.10 | -0.41 | 4th | 18b | 1 | 1.12 | -0.31 | 4th |
| Q26 | Circumstances | 0.50 | 0.48 | 1st | 18b | 2-5 times | 1.00 | -0.54 | 4th |
| Q26 | Not appli | -0.79 | 0.92 | 2nd | 18b | 6-10 times | 1.03 | -0.29 | 4th |
| Q26 | Not sure | -0.80 | 0.61 | 2nd | 17b | 6-10 times | 0.57 | -0.34 | 4th |
| Q26 | Parent/Caregiver | 0.04 | -0.17 | 4th | 18c | 1 | 0.88 | -0.55 | 4th |
| Q26 | State | 0.18 | 0.51 | 1st | 18c | 2-5 times | 1.07 | -0.56 | 4th |
| Q26 | Driver | 0.05 | -0.30 | 4th | 18c | 6-10 times | 1.12 | -0.56 | 4th |
| Q36 | Large city | 0.32 | -0.36 | 4th | 18d | > 10 times | 1.00 | -0.20 | 4th |
| Q36 | Medium sized city | -0.26 | 0.41 | 2nd | 18d | 1 | 0.97 | -0.52 | 4th |
| Q36 | Not appli | -1.08 | 1.14 | 2nd | 18d | 2-5 times | 1.12 | -0.72 | 4th |
| Q36 | Rural | 0.36 | 0.16 | 1st | 18d | 6-10 times | 1.03 | -0.71 | 4th |
| Q36 | Small town | 0.31 | 0.18 | 1st | 28c | Not at all important | 0.66 | -0.46 | 4th |
| Q36 | Suburb | -0.35 | -0.11 | 3rd | 28c | Some important | 0.31 | -0.28 | 4th |
| Q37 | 18-21 | 0.62 | 0.26 | 1st | 28c | Some unimportant | 0.85 | -0.37 | 4th |
| Q37 | 22-30 | 0.14 | -0.21 | 4th | Q24 | Never used taxis with kids | 0.32 | -0.15 | 4th |
| Q37 | 31-45 | -0.13 | 0.11 | 2nd | 17c | 1 | 0.26 | -0.19 | 4th |
| Q37 | 46-65 | -0.26 | 0.31 | 2nd | Q24 | Not at all confident | 0.37 | -0.32 | 4th |
| Q37 | Not appli | -1.08 | 1.14 | 2nd | Q24 | Not very confident | 0.87 | -0.41 | 4th |
| Q37 | Over 65 | -1.08 | 1.14 | 2nd | Q24 | Some confident | 0.84 | -0.45 | 4th |
| Q38 | Advanced degree | 0.19 | -0.14 | 4th | Q24 | Very confident | 0.54 | -0.19 | 4th |
| Q38 | College degree | -0.09 | -0.13 | 3rd | Q25 | Depends on the circumstances | 1.12 | -0.32 | 4th |
| Q38 | High school | 0.26 | 0.22 | 1st | 17c | 2-5 times | 0.58 | -0.45 | 4th |
| Q38 | Less than high school | -0.08 | 0.71 | 2nd | Q26 | Circumstances | 0.97 | -0.60 | 4th |
| Q38 | Not appli | -0.94 | 1.13 | 2nd | 17c | 6-10 times | 0.67 | -0.37 | 4th |
| Q38 | Some college | -0.07 | 0.16 | 2nd | | | | | |

*Note: Ques.= Question, Quand.= Quadrant, appli.=applicable.*

## Clusters for Riders used Uber with Kids

Figure 5 shows the TCA plot from the selected responses of Group 1 (have used RSSs with kids). There are three main clusters that show the most common answers from Group 1 for various questions. Cluster 1 includes participants who answered that they have high school, some college, or less than high school education. It also has participants that are located in a small town or rural area. It also has the answer of 1 time or 2-5 times for several parts of Question 17 and Question 18. Cluster 2 includes participants with a college degree and participants that are located in a suburb. The cluster includes answers that the driver or the parent/caregiver is responsible for child safety in a rideshare car (Question 25) and in a taxi (Question 26). Cluster 3 includes participants with an advanced degree who live in a large city. This cluster has the answer “very confident” that they followed child safety laws when using taxicab services. The cluster also includes the answer of 6-10 times or greater than 10 times for Question 17 and Question 18. The other two blue dotted boxes indicate that the similarity patterns of the responses are ‘not application’ or ‘zero.’

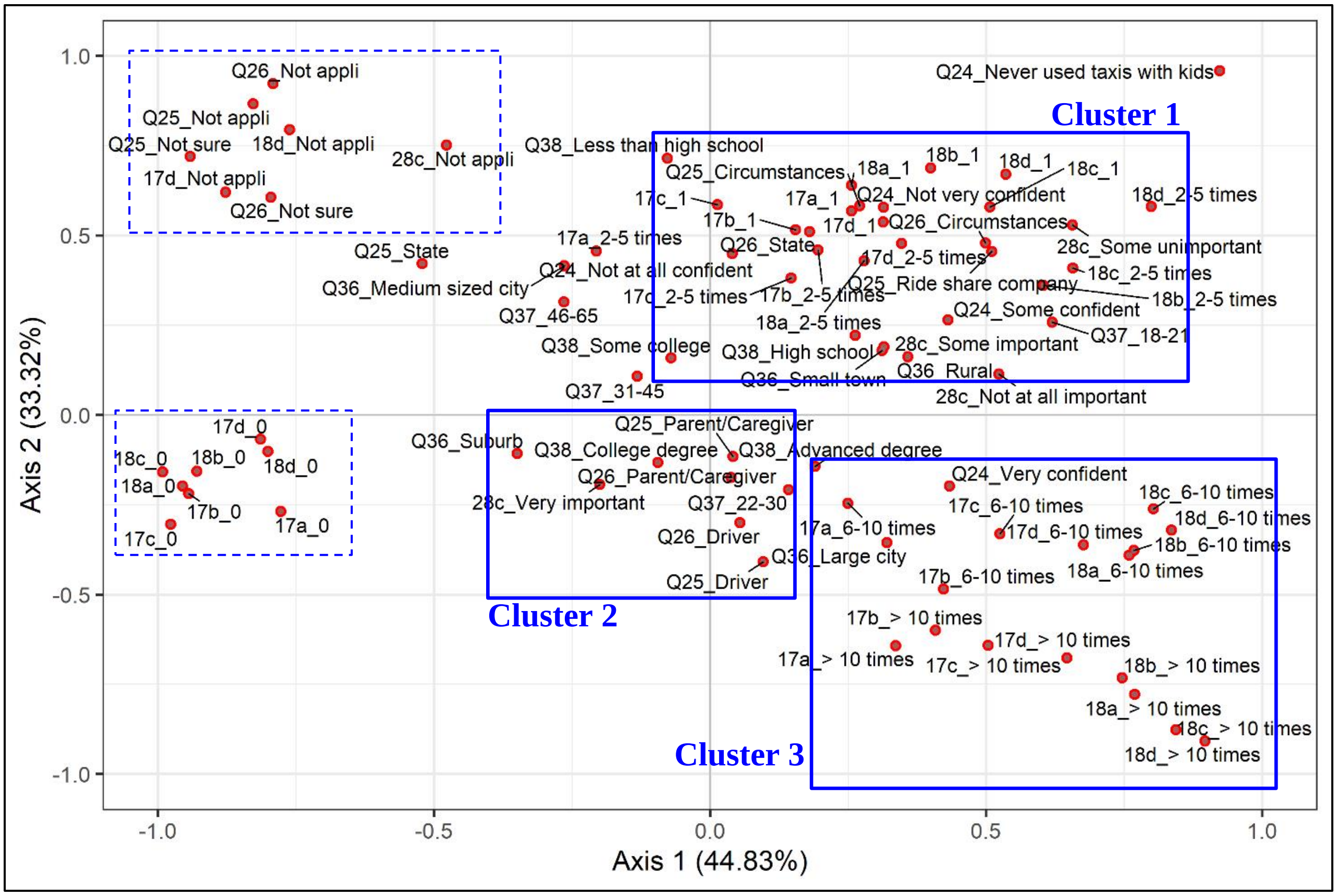

**Figure 5. TCA plot from selected responses of Group 1 (rideshared with kids).**

**Clusters for Riders never used Uber with Kids**

Figure 6 shows a TCA plot from the selected responses of Group 2 (never used RSSs with kids). There are three main clusters for Group 2 as well. Cluster 4 contains participants that answered to having high school or some college education and participants that are located in a small town, large city, or rural area. This cluster also includes the answer that the parent/caregiver is responsible for child safety in a rideshare car (Question 25) and in a taxi (Question 26). Cluster 5 includes participants that answered to having a college degree, advanced degree, or less than high school education and participants that are located in a suburb. This cluster also includes the answer that the driver is responsible, the rideshare company is responsible, or that they are not sure who is responsible for child safety in a rideshare car (Question 25) and in a taxi (Question 26). Cluster 6 includes answers that they are very confident, not very confident, or somewhat confident that they followed state laws while using taxi services with children. The dotted boxes shown in Figure 6 indicate the associations of some responses that are either ‘not applicable’ or ‘zero.’ The clusters show that less education and less familiarity with the ridesharing technology make people less confidence toward using ridesharing services.

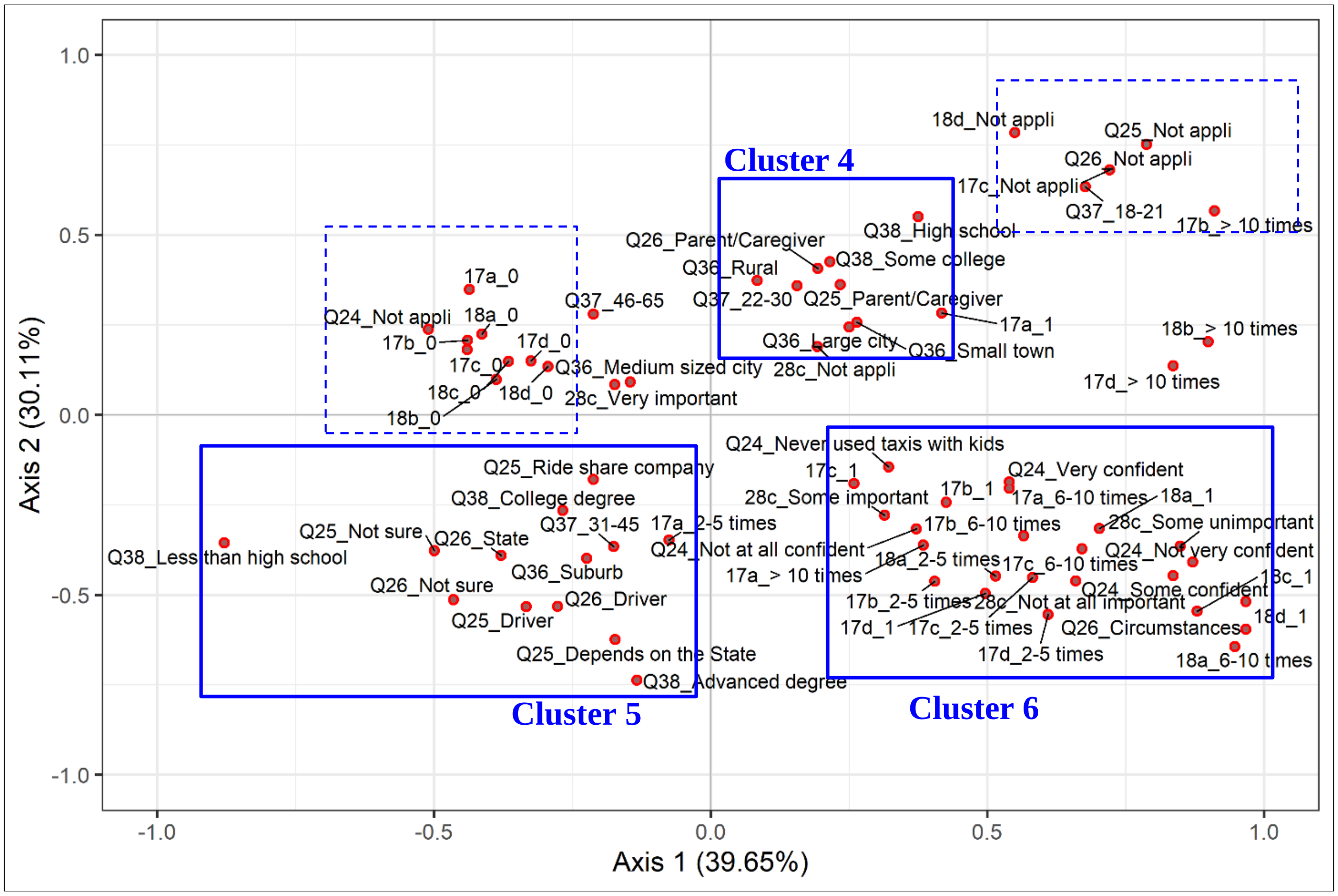

**Figure 6. TCA plot from selected responses of Group 2 (never rideshared with kids).**

## CONCLUSIONS

Serious and fatal injuries can be reduced significantly by seating young children in an appropriate position and securing them in car seats, booster seats, or seat belts that are age and size appropriate. CSSs are designed to promote child safety and protect children from being injured in car crashes. The improper use of CSSs is a pervasive and long-standing problem that negatively impacts the risk of injury and death in traffic crash occurrences. However, the usage of CSSs in RSSs is an area of research that is very limited, and there are very few studies focused on this important issue. The present study shows that there is significant disagreement among parents about child transportation safety in rideshare vehicles. It is most likely that these issues prevail due to the inconsistent regulations in the U.S., where it is the responsibility of each individual state to develop and communicate its own child passenger safety regulations. These state regulations are often difficult for riders and drivers of rideshare vehicles to find and comprehend; they sometimes involve multiple statutes, have confusing wording, or they can be overall vague and unclear. The current study highlights this important issue and analyzes a comprehensive survey data.

By performing a comparison test, this study found that response patterns vary widely among the participants of the two groups. The MCA analysis determined several patterns. It is found that urban-dwelling parents with higher education degrees eventually use RSSs often due to their familiarity with the ridesharing technology. They are well adapted to the use of technology-savvy RSSs with kids when necessary. On the other hand, non-urban and moderately educated parents and guardians are dismissive in using technology-savvy RSSs while having kids with them to ride. These opposite norms indicate the mental and psychological distinction between two groups. It is possible that urban-dwelling parents, although aware of the associated risks, often use RSSs without CSSs to save time and reduce complexity. There is a need for clear guidance to mandate CSSs in RSSs to improve child safety in traveling vehicles.

The current study is not without limitations. The current study comprises only two main research questions. Additional issues and insights can be drawn from this important survey. For example, personal traits can also be linked to the association between these two groups and their response patterns. Another caveat is that the current study is designed based on the survey responses of the participants by assuming that the responses of the participants reflect their real-life behaviors.

## ACKNOWLEDGEMENT

The author likes to thank SafeD UTC program and SafeD Dataverse (https://dataverse.vtti.vt.edu/dataverse/safed) for open sourcing this important survey data.

## REFERENCES

Allard, J., Choulakian, V., 2019. TaxicabCA: Taxicab Correspondence Analysis. https://cran.r-project.org/web/packages/TaxicabCA/TaxicabCA.pdf Access on August 1, 2020.

Asbridge, M., Ogilvie, R., Wilson, M., Hayden, J., 2018. The Impact of Booster Seat Use on Child Injury and Mortality: Systematic Review and Meta-Analysis of Observational Studies of Booster Seat Effectiveness. Accident Analysis & Prevention. 119, 50-57.

Billie, H., Crump, C.E., Letourneau, R.J., West, B.A., 2016. Child Safety and Booster Seat Use in Five Tribal Communities, 2010–2014. Journal of Safety Research. 59, 113-117.

Bromfield, N., Mahmoud, M., 2017. An Exploratory Investigation of Child Safety Seat Use Among Citizens of the United Arab Emirates. Journal of Transportation Safety & Security. 9(sup1), 130-148.

Brown, S.H., Grondin, D.E., Potvin, J.R., 2009. Strength Limitations to Proper Child Safety Seat Installation: Implications for Child Safety. Applied Ergonomics. 40(4), 617-621.

Choulakian, V., 2013. Graph Partitioning by Correspondence Analysis and Taxicab Correspondence Analysis. Journal of Classification. 30, 397-427.

Choulakian, V.,2006. L1-norm projection pursuit principal component analysis. Computational Statistics & Data Analysis. 50,1441 – 1451.

Choulakian, V., 2006. Taxicab correspondence analysis. Psychometrika. 71, 1-13.

Das, S., Sun, X., 2015. Factor Association with Multiple Correspondence Analysis in Vehicle–Pedestrian Crashes. Transportation Research Record: Journal of the Transportation Research Board. 2519(1), 95–103.

Das, S., Sun, X., 2016. Association Knowledge for Fatal Run-off-Road Crashes by Multiple Correspondence Analysis. IATSS Research. 39(2), 146–155.

Desapriya, E., Fujiwara, T., Scime, G., Babul, S., Pike, I., 2008. Compulsory Child Restraint Seat Law and Motor Vehicle Child Occupant Deaths and Injuries in Japan 1994-2005. International Journal of Injury Control and Safety Promotion. 15(2), 93-97.

Farmer, P., Howard, A., Rothman, L., Macpherson, A., 2009. Booster Seat Laws and Child Fatalities: A Case–Control Study. Injury Prevention. 15(5), 348-350.

Jalayer, M., Pour-Rouholamin, M., Zhou, H., 2018. Wrong-Way Driving Crashes: A Multiple Correspondence Approach to Identify Contributing Factors. Traffic Injury Prevention. 19(1), 35-41.

Jeffrey, J., Whelan, J., Pirouz, D. M., Snowdon, A. W., 2016. Boosting Safety Behaviour: Descriptive Norms Encourage Child Booster Seat Usage amongst Low Involvement Parents. Accident Analysis & Prevention. 92, 184-188.

Kakefuda, I., Yamanaka, T, Stallones, L., Motomura, Y., Nishida, Y., 2008. Child Restraint Seat Use Behavior and Attitude among Japanese Mothers. Accident Analysis & Prevention. 40(3), 1234-1243.

Kroeker, A.M., Teddy, A.J. Macy, M.L., 2015. Car Seat Inspection among Children Older Than 3 Years: Using Data to Drive Practice in Child Passenger Safety. Journal of Trauma and Acute Care Surgery. 79(3), S48-S54.

Kulanthayan, S., Razak, A., Schenk, E., 2010.Driver Characteristics Associated with Child Safety Seat Usage in Malaysia: A Cross-Sectional Study. Accident Analysis & Prevention. 42(2), 509-514.

Ma, S., Tran, N., Klyavin, V.E., Zambon, F., Hatcher, K.W., Hyder, A.A., 2012. Seat Belt and Child Seat Use in Lipetskaya Oblast, Russia: Frequencies, Attitudes, and Perceptions. Traffic Injury Prevention. 13(sup1), 76-81.

Moradi, M., Khanjani, N., Nabipour, A.R., 2017. An Observational Study of Child Safety Seat Use in an International Safe Community: Tehran, Iran. Traffic Injury Prevention. 18(1), 88-94.

Moradi, M., Khanjani, N., Nabipour, A.R., 2019. Determinants of Child Safety Seat Use among Parents in an International Safe Community, Tehran, Iran. Traffic Injury Prevention. 20(8), 844-848.

Nazif-Munoz, J. I., Nikolic, N., 2018. The Effectiveness of Child Restraint and Seat Belt Legislation in Reducing Child Injuries: The Case of Serbia. Traffic Injury Prevention. 19(sup1), S7-S14.

Ojo, T. K., 2018. Seat Belt and Child Restraint Use in A Developing Country Metropolitan City. Accident Analysis & Prevention. 113, 325-329.

Olsen, C.S., Cook, L.J., Keenan, H.T., Olson, L.M., 2010.Driver Seat Belt Use Indicates Decreased Risk for Child Passengers in a Motor Vehicle Crash. Accident Analysis & Prevention. 42(2), 771-777.

Owens, J. M., Womack, K. N., Barowski, L., 2019. Factors Surrounding Child Seat Usage in Rideshare Services. Office of the Assistant Secretary for Research and Technology.

Salvador, I., 2020. R Package 'compareGroups'. https://cran.r-project.org/web/packages/compareGroups/compareGroups.pdf Accessed August 1, 2020.

Smith, A., 2016. Shared, collaborative and on-demand: the new digital economy. Pew Research Center.

Sun, K., Bauer, M.J., Hardman, S., 2010. Effects of Upgraded Child Restraint Law Designed to Increase Booster Seat Use in New York. Pediatrics. 126(3), 484-489.

Thoreson, S., Myers, L., Goss, C., DiGuiseppi, C., 2009.Effects of a Booster Seat Education and Distribution Program in Child Care Centers on Child Restraint Use Among Children Aged 4 to 8 Years. Archives of Pediatrics & Adolescent Medicine. 163(3), 261-267.